\documentclass{aa}  

\usepackage{graphicx}

\usepackage[varg]{txfonts}
\usepackage{natbib,twoopt}
\usepackage{amsmath}
\usepackage[nopatch]{microtype}
\usepackage{booktabs}
\usepackage{multirow}
\usepackage{hyphenat}
\usepackage{ulem}
\usepackage{amssymb}
\usepackage{threeparttable}
\usepackage{xcolor}

\usepackage[colorlinks=True,allcolors=blue]{hyperref}

\def\asec{\ifmmode ^{\prime\prime}\else$^{\prime\prime}$\fi}
\def\msun{M$_{\odot}$}
\def\rsun{R$_{\odot}$}

\def\degs{\ifmmode ^{\circ}\else$^{\circ}$\fi}
\def\amin{\ifmmode ^{\prime}\else$^{\prime}$\fi}
\def\asec{\ifmmode ^{\prime\prime}\else$^{\prime\prime}$\fi}
\def\fss{\hbox{$.\!\!^{\rm s}$}}        
\def\farcs{\hbox{$.\!\!^{\prime\prime}$}}  
\def\farcm{\hbox{$.\!\!^{\prime}$}}  
\def\h{$^{\rm h}$}
\def\m{$^{\rm m}$}

\def\psr{J2017}
\def\psrn{J1513}
\def\hip{HiPERCAM}
\def\gaia{\textit{Gaia}}
\def\xmm{\textit{XMM-Newton}}

\def\magel{\textit{Magellan}}
\def\ergs{erg~s$^{-1}$}
\def\kirr{erg~cm$^{-2}$~s$^{-1}$~sr$^{-1}$}

\def\ps{Pan-STARRS}
\def\fermi{\textit{Fermi}}

\newcommand{\flux}{erg~s$^{-1}$~cm$^{-2}$}

\begin{document}
\titlerunning{PSR J1513$-$2550 and PSR J2017$-$1614} 
\authorrunning{Bobakov et al.}

   \title{Two black widow pulsars in the optical and X-rays}
   \author{A.V. Bobakov\thanks{bobakovalex@gmail.com}
          \inst{1},
           A. Yu. Kirichenko\inst{2,1},  
          S. V. Zharikov\inst{2}, 
           A. V. Karpova\inst{1},
           D. A. Zyuzin\inst{1},\\
           Yu. A. Shibanov\inst{1},  
            R. E. Mennickent\inst{3}
             \and
            D. Garcia-{\'A}lvarez\inst{4,5}
          }
   \institute{ Ioffe Institute, 26 Politekhnicheskaya, St. Petersburg, 194021,  Russia 
              \and           
               Instituto de Astronom\'ia, Universidad Nacional Aut\'onoma de M\'exico, Apdo. Postal 877, Baja California, M\'exico, 22800               
              \and
               Departamento de Astronom\'{i}a, Universidad de Concepci\'on, Casilla 160-C, Concepci\'{o}n, Chile
              \and
               Instituto de Astrof\'isica de Canarias, V\'{i}a L\'actea s/n, E-38200 La Laguna, Tenerife, Spain
              \and
               GRANTECAN, Cuesta de San Jos\'e s/n, E-38712 Bre\~na Baja, La Palma, Spain 
             }

   \date{Received ..., 2024; accepted ...}

   \abstract
   {
    Two millisecond  pulsars, PSR J1513$-$2550 and PSR J2017$-$1614, with spin periods of about 2.1 and 2.3  ms were recently discovered in the radio and $\gamma$-rays and classified as black widow pulsars in tight binary stellar systems with orbital periods of about 4.3 and 2.3 h.  
   }{
    Our goals are to reveal the fundamental parameters of both systems and their binary components using multi-wavelength observations.    
   }{
    We carried out the first time-series multi-band optical photometry of the objects with the 2.1-metre telescope of the Observatorio Astron\'omico Nacional San Pedro M\'artir, the 6.5-metre \magel-1 telescope, and the 10.4-metre Gran Telescopio Canarias. To derive the parameters of both systems, we fitted the obtained light curves with a model assuming heating of the companion by the pulsar. We also analysed  archival  X-ray data  obtained with  the \xmm\ observatory. 
   }{
    For the first time, we firmly identified J1513$-$2550 in the optical and both pulsars in X-rays. The optical light curves of both systems have a single peak per orbital period with a peak-to-peak amplitude of $\gtrsim2$ magnitudes. The J2017$-$1614 light curves are symmetric, while  J1513$-$2550 demonstrates strong asymmetry whose nature remains unclear.
   }{
    We constrained the orbital inclinations, pulsar masses, companion temperatures and masses, and the distances to both systems. We also conclude that J2017$-$1614 may contain a massive neutron star of 2.4$\pm$0.6 \msun. The X-ray spectra of both sources can be fitted by  power laws with parameters typical of black widow systems.
   }

   \keywords{stars: neutron, stars: binaries: close, pulsars: individual (PSR J2017$-$1614, PSR J1513$-$2550)}
   \maketitle
  
   \section{Introduction}
        \label{sec:introduction}
        
        \begin{table*}[hbt!]
            \begin{center}
                \begin{threeparttable}
                    \caption{\psrn\ and \psr\ parameters obtained from \citet{sanpaarsa},  the ATNF catalogue, and the Third \fermi\ Large Area Telescope 
                    catalogue of $\gamma$-ray pulsars \citep{III-fermi}.}
                    \label{tab:pars}
                    \begin{tabular}{lcc}
                        \hline
                        MSP                                               & \psr\                           & \psrn\                            \\
                        \hline                                                                                                                    
                        R.A. $\alpha$ (J2000)                             & 20\h17\m46\fss1478(8)           & 15\h13\m23\fss32059(6)            \\
                        Dec. $\delta$ (J2000)                             & $-$16\degs14\amin15\farcs51(5)  & $-$25\degs50\amin31\farcs285(3)   \\
                        Galactic longitude $l$, deg                       & 27.314                          & 338.820                           \\
                        Galactic latitude $b$, deg                        & $-$26.220                       & 26.964                            \\
                        Spin period $P$, ms                               & 2.3142872649224(4)              & 2.1190675651177(1)                \\
                        Period derivative $\dot{P}$, s s$^{-1}$           & 2.45(5)$\times$10$^{-21}$       & 21.61(2)$\times$10$^{-21}$        \\
                        Orbital period $P_b$, d                           & 0.0978252578(4)                 & 0.1786354505(8)                   \\
                        Projected semi-major axis $x$, lt-s               & 0.043655(5)                     & 0.0408132(7)                      \\
                        Epoch of ascending node $T_{\rm asc}$, MJD        & 56704.756314(2)                 & 56728.4539340(6)                  \\
                        Mass function $f_M$, \msun\                       & 9.334(3)$\times$10$^{-6}$       & 2.2874(1)$\times$10$^{-6}$        \\
                        Dispersion measure DM, pc~cm$^{-3}$               & 25.4380(4)                      & 46.86(7)                          \\
                        Distance $D_{\rm YMW16}$, kpc                     & 1.4                             & 4.0                               \\
                        Distance $D_{\rm NE2001}$, kpc                    & 1.1                             & 2.0                               \\
                        Characteristic age $\tau_c\equiv P/2\dot{P}$, Gyr & 15                              & 1.55                              \\
                        Spin-down luminosity $\dot{E}$, erg s$^{-1}$      & 7.8$\times$10$^{33}$            & 9.0$\times$10$^{34}$              \\
                        Minimum companion mass $M_{\rm c,\ min}$, \msun   & 0.03                            & 0.02                              \\
                        \hline                                                                                                                    
                        p.m. in R.A. direction $\mu_\alpha$cos$\delta$, mas yr$^{-1}$           & --        &      $-$6.3(1)                    \\
                        p.m. in Dec. direction $\mu_\delta$, mas yr$^{-1}$       & -- &      $-$3.5(4)\\                                          
                        \hline
                    \end{tabular}
                    \begin{tablenotes}[hang]
                        \item Numbers in parentheses denote 1$\sigma$ uncertainties relating to the last 
                        significant digit.
                        \item $D_{\rm YMW16}$ and $D_{\rm NE2001}$ are 
                        dispersion measure distances derived using the YMW16 \citep*{ymw2016} 
                        and NE2001 \citep{ne2001} models for the distribution of free electrons 
                        in the Galaxy, respectively.
                        \item The minimum companion mass was calculated for the pulsar 
                        mass of $M_{\rm p}=1.4~$\msun\  and orbital inclination of $i=90$\degs.
                        \item p.m. = proper motion.
                    \end{tablenotes}
                \end{threeparttable}
            \end{center}
        \end{table*}

        To date, there are over 550  pulsars\footnote{According to the ATNF pulsar catalogue \citep{atnf}; \url{https://www.atnf.csiro.au/people/pulsar/psrcat/}}  with spin periods $\lesssim30$ ms. Such rapidly rotating neutron stars (NSs) are named millisecond pulsars (MSPs). The commonly accepted `recycling' scenario of their   formation \citep{Bisnovatyi-Kogan1974,alpar1982} suggests that these NSs are spun up in binary systems through accretion from a secondary star, which eventually becomes a low-mass white dwarf \citep{tauris2011}. This also explains why most MSPs are found in binary systems. However, not all binary companions to MSPs are white dwarfs: some of them tend to have a more exotic nature. Those binary MSPs that contain low-mass non- or semi-degenerate companions in tight orbits with periods of $P_b\lesssim1$ d are called `spider' pulsars. This class includes redbacks (RBs) and black widows (BWs) \citep[see, e.g.][]{roberts2013}. In these systems, one side of a spider companion is irradiated by the energetic pulsar wind, which leads to  heating and evaporation of its stellar material. The RBs have more massive companions with masses of 0.1--1~\msun,  while masses of BW companions are $\lesssim0.05$~\msun. The formation of such systems as well as the link between the two subclasses are matters of debate \citep{chen2013,benvenuto2014,ablimit2019,guo2022}.

        The number of known spiders has increased significantly in the last few years. Many of them were discovered thanks to the multi-wavelength studies of unidentified $\gamma$-ray sources detected with the \fermi\ observatory  \citep[e.g.][]{salvetti2017,swihart2022}. In addition to a broad range of spiders' fundamental parameters commonly derived from radio follow-up searches, optical observations can also have a significant impact. In particular, they allow one to derive the temperature distribution over a companion surface, its Roche-lobe filling factor,   irradiation efficiency, distance to the binary, and masses of its components \citep{strader2019,draghis2019,swihart2022}. With a larger sample of newly detected spider systems, further investigations can help to clarify their origin and evolution.

        Two BW pulsars, PSRs J2017$-$1614 and  J1513$-$2550  (hereafter \psr\ and \psrn),  were discovered in the radio with the Green Bank Telescope  during the search for counterparts of unassociated  \fermi\ sources \citep{sanpaarsa}. Their parameters are listed in Table~\ref{tab:pars}. The radio emission of both binary MSPs was eclipsed, but only in some observational runs. After the discovery in the radio, their periodic pulsations were also found in $\gamma$-rays \citep{sanpaarsa}, but no studies in X-rays have been reported.

        The optical counterpart of \psr\ was first observed in the $R$ band with the 2.4-metre telescope of the Michigan-Dartmouth-MIT (MDM) observatory \citep{sanpaarsa}. It shows significant magnitude  variations, from $R\approx21.8$ mag at the maximum  brightness to $R\gtrsim24$ mag at the minimum, with a period of $\approx$2.3 h, which is compatible with the binary period derived from the radio data \citep{sanpaarsa}.

        A possible  optical companion of \psrn\ could be preliminarily identified by positional coincidence  with  the star PSO  J228.3472$-$25.8420 from the Panoramic Survey Telescope and Rapid Response System DR 2 catalogue (\ps, \citealt{flewelling2020}). Its  mean brightness in the $i$-band is about 22 mag.

        No detailed optical studies of \psrn\ and \psr\ have been reported so far. Here, we present the results of the first multi-band time-series optical  photometry of \psr\ and \psrn\  and analyse X-ray data for both BWs obtained with the \xmm\ observatory. Optical observations and data reduction are described in Sec.~\ref{sec:data} and analysed in Sec.~\ref{sec:lc}, X-ray data are presented in Sec.~\ref{sec:xray}, and the results are discussed and concluded in Sec.~\ref{sec:discussion} and \ref{sec:conclusion}.

        \begin{table*}
            \caption{Log of the \psrn\ and \psr\ observations.}
            \begin{center}
                \begin{tabular}{ccccccc}
                    \hline
                    MJD&  Date&  Filter&  Exposure time, s &  Airmass& Seeing, arcsec & Binning\\
                    \hline    
                    \multicolumn{7}{c}{\textbf{\psr\ / GTC} } \\ 
                        59464.91             & 2021-09-07  & $u_{s}$ & 79$\times$140 & 1.41--1.83 & 0.6--1.3 & 1$\times$1\\ 
                                             &             & $g_{s}$ & 79$\times$140 &            &          & 1$\times$1\\
                                             &             & $r_{s}$ & 79$\times$140 &            &          & 1$\times$1\\
                                             &             & $i_{s}$ & 79$\times$140 &            &          & 1$\times$1\\
                                             &             & $z_{s}$ & 79$\times$140 &            &          & 1$\times$1\\
                    \hline
                    \multicolumn{7}{c}{\textbf{\psrn\ / OAN-SPM 2.1-m} } \\ 
                        58604.81             & 2019-05-01  &   $R$   & 17$\times$600 & 1.9--2.2   &  1.9--2.4 & 2$\times$2\\
                        58605.78             & 2019-05-02  &   $R$   & 22$\times$600 & 2.0--2.4   &  1.7--1.8 & 2$\times$2\\ 
                        58606.78             & 2019-05-03  &   $B$   & 5$\times$600  & 2.0--2.4   &  1.9--2.6 & 2$\times$2\\    
                                             &             &   $V$   & 5$\times$600  &            &           & 2$\times$2\\ 
                                             &             &   $R$   & 5$\times$600  &            &           & 2$\times$2 \\ 
                                             &             &   $I$   & 5$\times$600  &            &           & 2$\times$2 \\ \\
                    \multicolumn{7}{c}{\textbf{\psrn\ / \magel} } \\
                        59692.23             & 2022-04-23  &   $r'$   & 64$\times$50  & 1.0--1.2   & 0.4--0.6 & 1$\times$1 \\ 
                        59692.34             & 2022-04-23  &   $r'$   & 57$\times$50 & 1.2--1.7   & 0.4--0.5 & 2$\times2$\\  
                    
                    \hline      
                \end{tabular}
            \end{center}
            \label{tab:ObservationLog}
        \end{table*}

    \section{Optical observations, data reduction, and calibration}
        \label{sec:data} 
        \begin{figure*}[t]
            \begin{center}
                \includegraphics[width=10.37cm, clip]{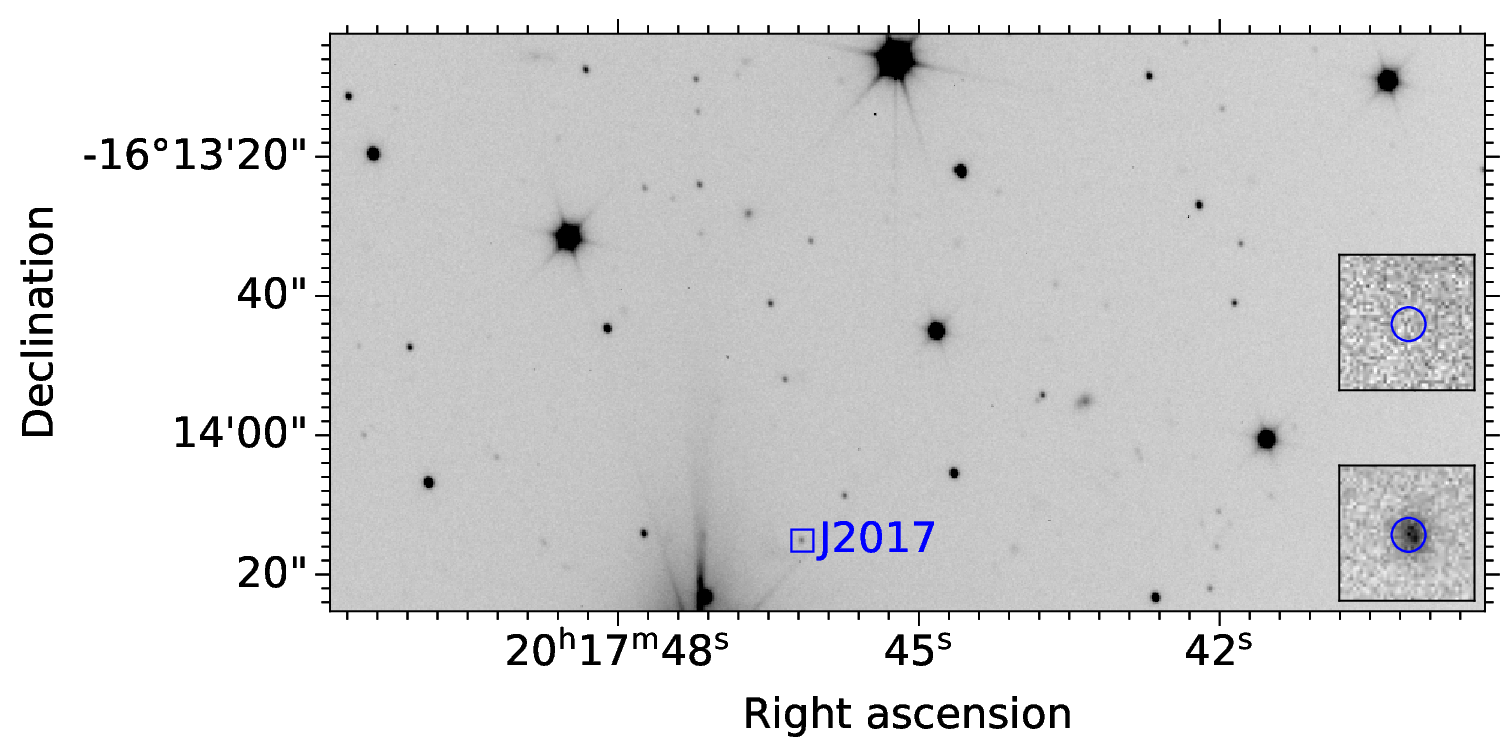}
                \includegraphics[width=6.46cm, clip]{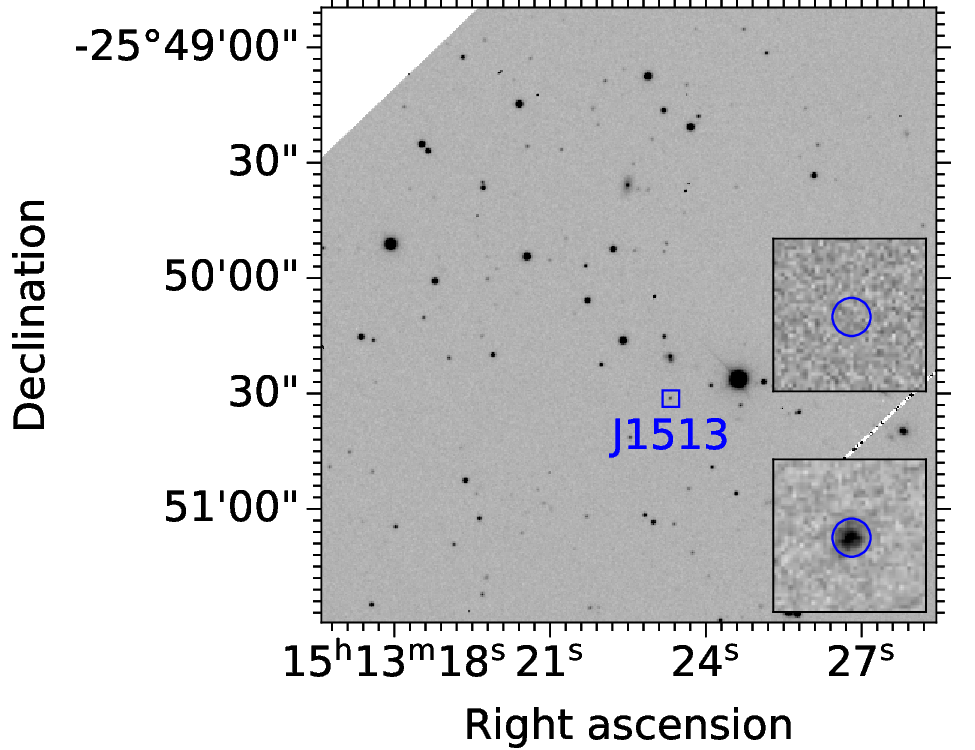}    
            \end{center}
            \caption {
                Optical images of the fields of the two pulsars.
                \textit{Left panel}: $2\farcm8 \times 1\farcm4$ FoV of the \hip\ instrument in the $r_s$ band containing \psr. 
                \textit{Right panel}: $2\farcm7 \times 2\farcm7$ image of the \psrn\ field in the $r'$ band obtained with the \magel-1 telescope. The pulsars' vicinities within the small  boxes are enlarged in the insets to demonstrate the maximum (bottom) and minimum (top) brightness phases of their companions, the positions of which are marked by blue circles.
            }
            \label{fig:img}
        \end{figure*}

        \subsection{\psr}
        
            The phase-resolved photometric observations of the \psr\ field\footnote{Programme GTC8-21AMEX, PI A. Kirichenko} were performed on September 7 2021 with the \hip\ instrument\footnote{\url{http://www.gtc.iac.es/instruments/hipercam/hipercam.php}} \citep{dhillon2016,dhillon2018,dhillon2021} at the 10.4-metre Gran Telescopio Canarias (GTC). The \hip\ allows for simultaneous imaging in five ($u_s$, $g_s$, $r_s$, $i_s$, and $z_s$) high-throughput `Super' SDSS filters. 
            
            In each band, we obtained 79 images with individual exposure times of 140-s. The field of view (FoV) is $2\farcm8\times1\farcm4$, with the 0\farcs081 image scale in the 1$\times$1 binning mode. The weather conditions during the observations were photometric. The log of observations is presented in Table~\ref{tab:ObservationLog}. 
            
            We reduced the data following the \hip\ pipeline manual,\footnote{\url{https://cygnus.astro.warwick.ac.uk/phsaap/hipercam/docs/html/}} including bias subtraction and flat field correction  \citep{dhillon2016}. We also produced bad pixel correction with the defect files obtained from the \hip\ website. In addition, we applied the defringing procedure to the $z_s$-band images. An example of an individual image in the $r_s$-band is presented in Fig.~\ref{fig:img}, left panel. The pulsar counterpart position is marked by the blue box. Its variability is demonstrated in the insets.
            
            Using the optimal extraction algorithm \citep{optimal_slgorithm}, we measured instrumental magnitudes of the counterpart and a dozen of bright stars in the FoV whose magnitudes are available in the \ps\ catalogue. To avoid centroiding problems of the companion during its faint brightness stages, its position was fixed relative to a nearby star when the source was in its maximum brightness phase. We  performed photometric  calibration  using the GTC atmospheric extinction coefficients $k_{u_s}=0.48$, $k_{g_s}=0.17$, $k_{r_s}=0.1$, $k_{i_s}=0.05$, and $k_{z_s}=0.05$ and the spectro-photometric standards WD1606+422 and WD2047+372\footnote{\url{http://www.vikdhillon.staff.shef.ac.uk/hipercam/hcam_flux_stds.pdf}} \citep{dhillon2021} observed during the same night as the target. To eliminate possible systematic errors and to account for photometric zero-point variations during the observation, we used the technique described in \citet{LCcorrTechHonneycutt} and the \ps\ stars mentioned above. The measured magnitudes of the latter were compared with their catalogue values applying the photometric transformation equations, which establish  relations  between the magnitudes in the \hip\ and \ps\ bands   \citep{Pan_Strarrs_toSDSS}. This allowed us to verify the calibration, which resulted in the final zero-points $z_{u_{s}} = 27.30(5)$ mag, $z_{g_{s}} = 27.80(5)$ mag, $z_{r_{s}} = 27.65(5)$ mag, $z_{i_{s}} = 27.41(5)$ mag, and $z_{z_{s}} = 27.06(5)$ mag. The detection limits varied in the ranges of $u_s$ = 24.9--25.2 mag, $g_s$ = 23.5--25.0 mag, $r_s$ = 23.3--24.5 mag, $i_s$ = 22.7--24.2 mag, and $z_s$ = 23.0--24.3 mag.

            In the $u_s$-band, at given individual exposures the J2017 companion brightness was below   the \hip\ sensitivity limit.  As a result, we obtained \psr\ light curves only  in the $g_s$, $r_s$, $i_s$, and $z_s$ bands.

        \subsection{\psrn}
            Time-series optical photometry of \psrn\  was carried out using the 2.1-metre telescope of Observatorio Astron\'omico Nacional San Pedro M\'artir (OAN-SPM) and the 6.5-metre \magel-1 telescope (Baade). 
            
            OAN-SPM observations were performed in the $B$, $V$, $R$, and $I$ bands with  the `Rueda Italiana' instrument during three observing runs in May  2019. The detector FoV is $6\amin \times6\amin$ with a pixel scale of 0\farcs34 in the 2$\times$2 CCD pixel binning mode. \magel-1 observations were carried out in the Sloan $r'$ band   with the Inamori-Magellan Areal Camera and Spectrograph (IMACS) during one observing run in April 2022 (PI R. E. Mennickent). The pixel scale was 0\farcs111 in the 1$\times$1  CCD pixel binning mode. The target was exposed on chip 3 with the FoV of $3\farcm8 \times7\farcm5$. The weather conditions were photometric during both observing runs. The log of all observations is given in Table~\ref{tab:ObservationLog}.

            Standard data reduction, including bias subtraction, flat-fielding, and cosmic ray rejection, was performed  for each dataset utilising the Image Reduction and Analysis Facility (\texttt{IRAF}) package. The defringing procedure was applied to the OAN-SPM images in the $I$-band.

            Astrometric referencing was performed using 12 stars from the \gaia\ DR 2 astrometric catalogue \citep{gaia2016,gaia2018} Formal $rms$ uncertainties of the resulting astrometric fit were $\Delta\alpha$~$\lesssim0\farcs08$ and $\Delta\delta$~$\lesssim0\farcs08$ for the \magel-1 and $\Delta\alpha$~$\lesssim0\farcs04$ and $\Delta\delta$~$\lesssim0\farcs04$ for the OAN-SPM.

            An example of an individual 50~s $r'$-band image of the \psrn\ field obtained  with the \magel-1 telescope is shown in the right panel of Fig.~\ref{fig:img}. The co-ordinates of the \psrn\ detected companion candidate (marked by the blue circle in the inserts) are fully consistent with the pulsar radio position given in Table~\ref{tab:pars}, confirming  its preliminary identification based on the \ps\ catalogue. 
            
            Photometric calibrations were performed using photometric standards observed during the same nights as the target. For the \magel-1 data, we used three standards from the PG 1525-071 field \citep{1992AJ....104..340L} and the atmospheric extinction coefficient $k_{r'} = 0.1$ provided by the IMACS team.\footnote{\url{https://www.lco.cl/technical-documentation/imacs-user-manual/}} As a result, the zero point in the $r'$-band was 28.15(1) and the 3$\sigma$ sensitivity level was $\approx24.8$ in the AB system. For the  OAN-SPM data, we used the  PG1323 spectrophotometric standard and the extinction coefficients $k_{B} = 0.25$ mag, $k_{V} = 0.14$ mag, $k_{R} = 0.07$ mag, and $k_{I} = 0.06$ mag relevant to the telescope site. The resulting zero points were $z_R = 24.96(2)$ mag, $z_B = 24.91(3)$ mag, $z_V = 25.16(2)$ mag, and $z_I = 24.15(2)$ mag, and the 3$\sigma$ sensitivity levels were $\approx 23.6$ mag, $\approx 24.1$ mag, $\approx 23.8$ mag, and $\approx 22.0$ mag, respectively.
            \footnote{The $R$-band zero point remained the same during the three nights of observations in photometric conditions.}

            We performed standard aperture photometry of the source and a number of stars from the \ps\ catalogue in each image. Aperture radii were chosen according to the optimal extraction algorithm \citep{optimal_slgorithm}. The measured magnitudes of the stars from the catalogue remained consistent with their catalogue values during the observations.

    \section{Light curve analysis}
        \label{sec:lc}
        
        \begin{figure*}
        \centering
            \begin{center}
                \includegraphics[width=0.8\textwidth]{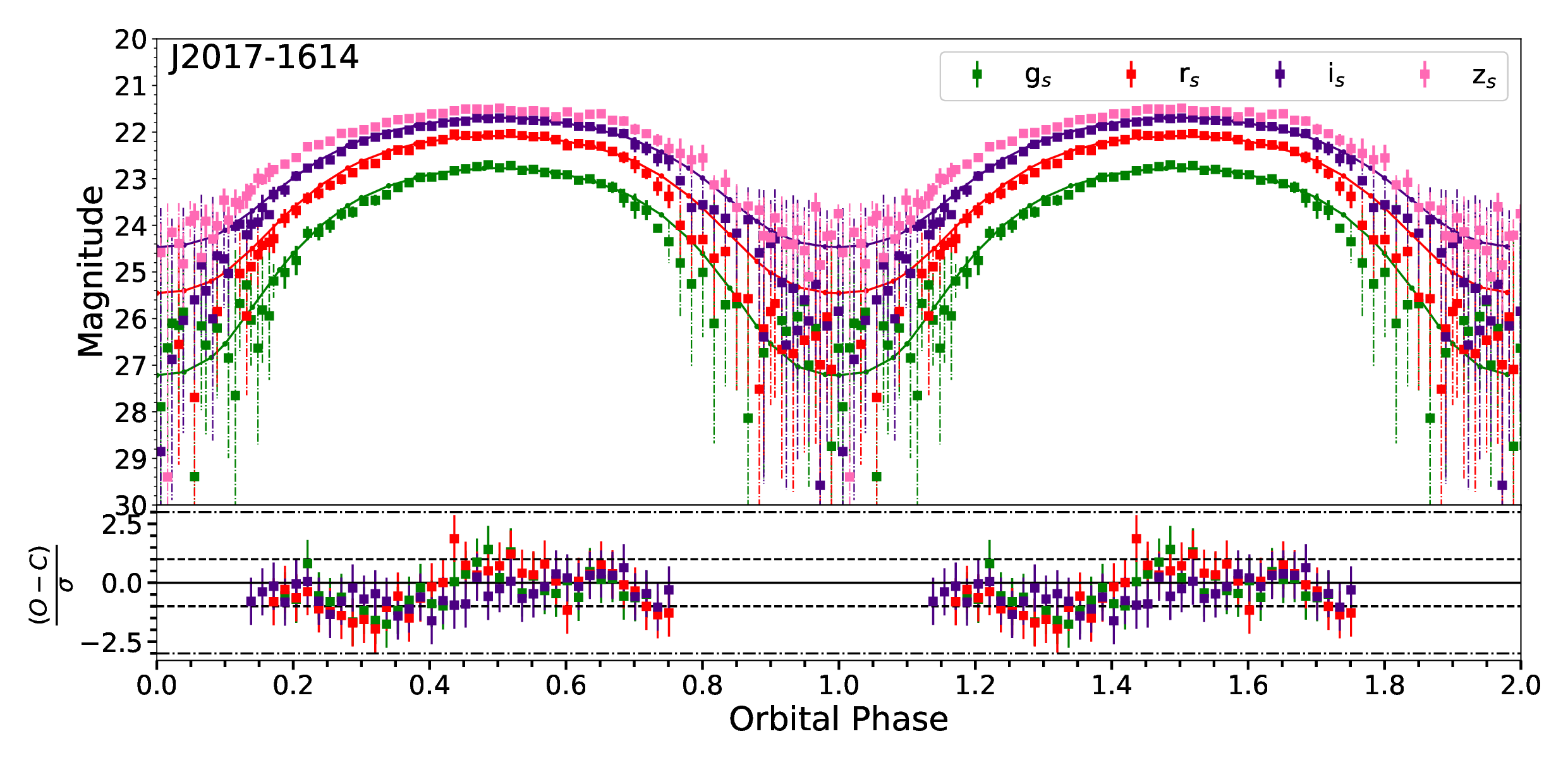} 
                \includegraphics[width=0.8\textwidth]{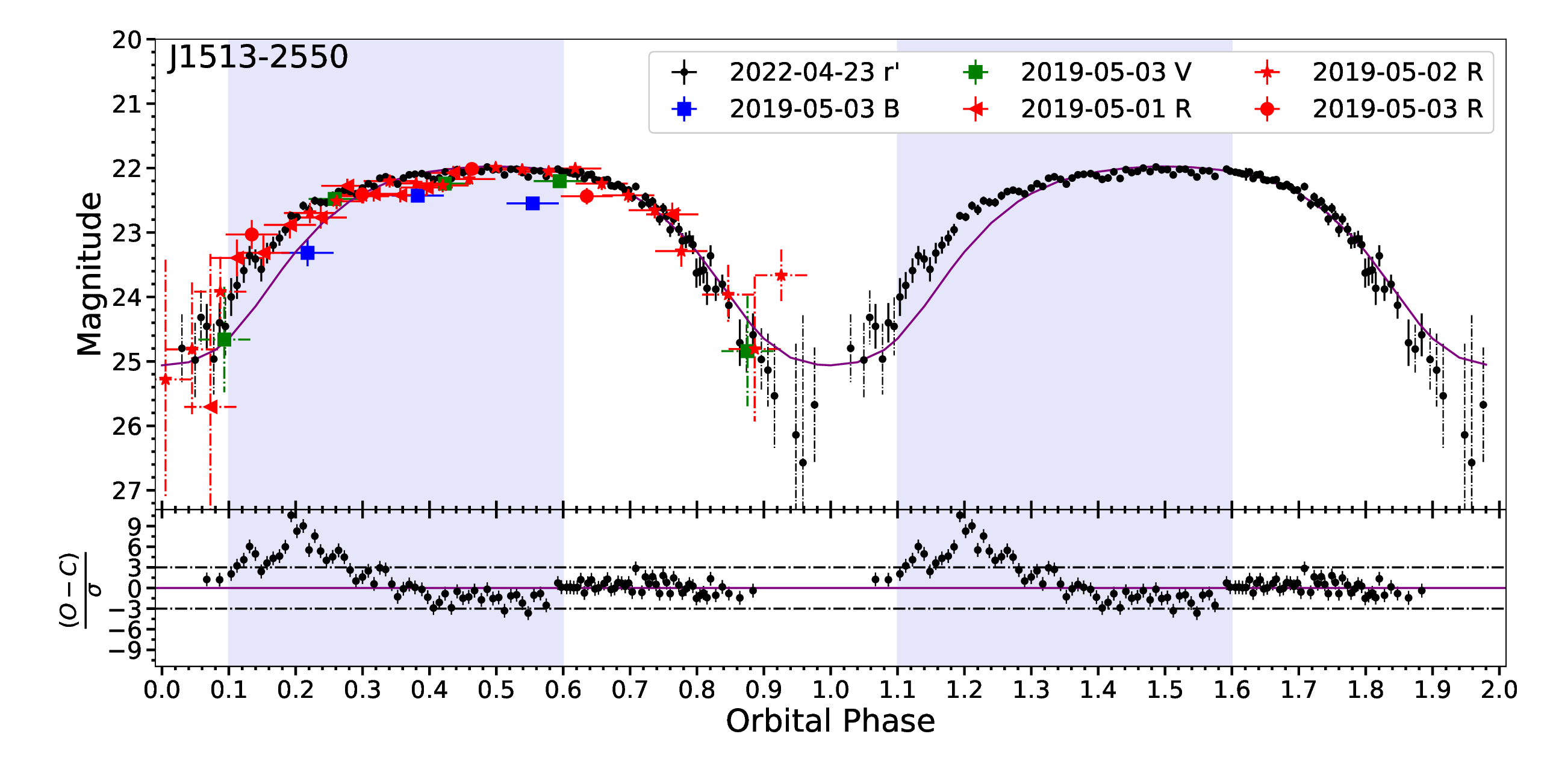}
            \end{center}
            \caption{
                   Light curves of the two pulsars and results of their modelling.  
                   \textit{Top panel}: $g_s$, $r_s$, $i_s$, and $z_s$ band light curves of the \psr\ system folded with the orbital period and the best-fitting model (solid lines). Two periods are shown for clarity. 
                   \textit{Bottom panel}: \psrn\ light curves folded with the orbital period. Data obtained in different filters and different nights are shown with various colours and symbols, indicated in the legend. All magnitudes are in the AB system. The best-fitting model is shown by the solid line. Shaded areas mark the part of the light curves where small-scale wavering was detected (see text for details). 
                   In both panels, points with dashed error bars lie below the 3$\sigma$ detection limits. The phases $\phi=0.0$ correspond to the minima of the systems' model brightness. Sub-panels show residuals calculated as the difference between the observed ($O$) and calculated ($C$) magnitudes for each data point in terms of the magnitude error, $\sigma$.
               }
             \label{fig:lc}
        \end{figure*}

        \begin{table}
            \caption{The light-curve fitting results for \psrn\ and \psr.}
            \label{tab:fit} 
            \begin{center}
                \begin{threeparttable}
                    \begin{tabular}{lcc}
                        \hline                                                                                                                 
                        Fitted parameters                                          & \psr                     & \psrn                    \\    
                        \hline                                                                                                                 
                        Reddening $E(B-V)$, mag                                    & 0.09$_{-0.01}^{+0.03}$   & 0.07$_{-0.04}^{+0.03}$   \\    
                        Distance $D$, kpc                                          & 2.40$_{-0.05}^{+0.10}$   & 1.95(5)                  \\    
                        Pulsar mass $M_{\rm p}$, \msun                             & 2.4(6)                   & 1.7$_{-0.6}^{+1.0}$      \\    
                        Mass ratio $q$ =  $M_{\rm c}/M_{\rm p}$                    & 0.017(2)                 & 0.012(1)                 \\    
                        `Night-side' temp. $T_{\rm n}$, 10$^3$ K                   & 3.0$_{-0.1}^{+0.2}$      & 3.2(1)                   \\    
                        Inclination $i$, deg                                       & 67$_{-7}^{+10}$          & 70$_{-15}^{+20}$         \\    
                        Roche lobe filling factor $f$                              & 0.84(6)                  & 1.0$_{-0.1}^{+0.0}$      \\    
                        Irradiation factor $K_{\rm irr}$,                          & \multirow{2}{*}{0.7(1)}  &  \multirow{2}{*}{0.6(1)} \\    
                          10$^{20}$ \kirr                                          &                          &                          \\    
                        \hline                                                                                                                 
                        Derived parameters                                         &                          &                          \\    
                        \hline                                                                                                                 
                        Companion mass $M_{\rm c}$, \msun                          & 0.041                    & 0.020                    \\    
                        Companion radius $R_{\rm c}^x$, \rsun                      & 0.122                    & 0.114                    \\    
                        Companion radius $R_{\rm c}^y$, \rsun                      & 0.169                    & 0.151                    \\    
                        Min `day-side' temp. $T_{\rm d}^{\rm min}$, 10$^3$ K       &  3.3                     & 3.2                      \\    
                        Max `day-side' temp. $T_{\rm d}^{\rm max}$, 10$^3$ K       & 5.9                      & 5.4                      \\    
                        Irradiation efficiency $\eta$                              & 0.6                      &  0.05                    \\    
                        \hline                                                                                                   
                    \end{tabular}
                    \begin{tablenotes}
                        \item Irradiation efficiencies $\eta$ are calculated using the spin-down luminosities from Table~\ref{tab:pars}.
                        \item $R_{\rm c}^x$ and $R_{\rm c}^y$ are the 
                        radii of the ellipsoidal companion. 
                        The latter is along the line passing through the centres of the binary sources.
                    \end{tablenotes}
                \end{threeparttable}
            \end{center}
        \end{table}
        
        The \psr\ and \psrn\ optical companions' barycentre-corrected light curves folded with their known orbital periods (see Table~\ref{tab:pars}) are presented in  Fig.~\ref{fig:lc}. The times of the ascending node were chosen as the reference times for each band. The curves show one broad peak per period and peak-to-peak amplitudes of $\gtrsim 2$ mag, as was expected for BW systems in which the heating of the companion by the pulsar is dominated over the ellipsoidal modulation \citep[e.g.][]{draghis2019,matasanchez2023}. This firmly establishes both optical sources as the true companions of the considered binary MSPs.

        The light curves of \psr\ are symmetric relative to the brightness maxima where the colour indices are $g_{s}-r_{s}=0.7$, $r_{s}-i_{s}=0.4$, and $i_{s}-z_{s}=0.2$. The colours slowly become redder as the system rotates towards the minimum, reflecting the difference between the `day-side' and `night-side' temperatures of the pulsar companion. When the system approaches the minimum, its brightness drops below the detection limit.

        The \psrn\ light curves are more complex. They demonstrate significant asymmetry with a steeper initial brightness increase and a more gradual  fading to the minimum. We also note  apparent regular brightness variations (wavering) on a short timescale with a peak-to-peak amplitude of $\sim  0.1$ mag, which are clearly seen in the $r'$-band at the rising part of the light curve within the  orbital phase range of 0.1--0.6. They are also present in the $R$-band curve, though with smaller significance due to the worse time resolution and sensitivity. The variations are not visible in the phase range of 0.6--1.0 in either of the two bands. To check whether they are caused by some atmospheric effects, we investigated light curves of several isolated stars with magnitudes within the range of $r=17.8$--20 mag in the \psrn\ field. We did not find similar patterns there. For this reason, we concluded that the observed features are real and can be attributed to some physical process in the \psrn\ system.
        
        To derive  parameters of the systems, we fitted the obtained light curves applying a binary model consisting of an NS and a low-mass companion heated by the pulsar wind. The model and fit method are  described in detail in \citet{zharikov2013,zharikov2019}. It is assumed that each surface element of the companion emits a black-body spectrum whose effective temperature varies from element to element. This direct heating model produces symmetrical light curves in respect to their brightness maxima.
        
        In the case of \psr, we used only $g_s$-, $r_s$-, and $i_s$-band light curves because the black-body  approximation does not describe well the observed spectral energy distribution in the wide spectral range from the $g_s$- to $z_s$ band. As for \psrn, since the utilised model does not include components that can describe the asymmetry and small-scale variations in the light curves, we fitted only the smooth part of the $r'$-band light curve within the orbital phase range of $\phi \in [0.6, 1.0]$. We did not include the other bands in the fitting procedure as their respective light curves contain too few data points, and some of them cover phases with variations. For both objects, we excluded points that lie below the 3$\sigma$ detection limits from the modelling.
        
        The fitted parameters are the interstellar reddening, $E(B-V)$, the distance, $D$, the pulsar mass, $M_{\rm p}$, the orbit inclination angle,~$i$, the effective irradiation factor, $K_{\rm irr}$, which defines heating of the companion, the companion Roche lobe filling factor, $f$, and  the companion `night-side' temperature, $T_{\rm n}$. The mass of the companion, $M_{\rm c}$, or the mass ratio, $q=M_{\rm c}/M_{\rm p}$, was defined by the mass function (Table~\ref{tab:pars}), which links $M_{\rm p}$, $M_{\rm c}$,  and $i$. The allowed range of $E(B-V)$ was limited by the total galactic interstellar absorption in the directions of the pulsars ($\approx$0.12 mag, \citealt{dustmap2019}). The masses of the pulsars started  from 1.4 M$_\odot$, and it was assumed that they cannot exceed  3 \msun. Other parameters did not have any special restrictions. They were started from random values inside the natural physical limitations. The gradient descent method was used to find the minimum of the $\chi^2$ function. The resulting parameters for both pulsars are presented in Table \ref{tab:fit} and the best-fitting models are shown by solid lines in Fig.~\ref{fig:lc}. 
        
        The light curves of the \psr\ are perfectly reproduced  by the model with reduced $\chi^{2} = 59.6$ for 99 degrees of freedom ({\sl dof}). The fit residuals are within the  2$\sigma$ data errors (Fig.~\ref{fig:lc}, top) and the fit parameters are consistent with those expected for BWs. In contrast, J1513 shows light curves different from most spider companions (see discussion below). Formally, the fit of a part of the light curve is acceptable, with  $\chi^{2}=69.4$ for 44 {\sl dof}. However,  as can be seen from the bottom panel of Fig.~\ref{fig:lc}, the observed light curve demonstrates a strong emission excess over the symmetric model within the range of $\phi\sim 0.1$--0.35, indicating the presence of some additional source of radiation at these orbital phases. Moreover, there are small-scale brightness variations mentioned above that deserve a separate analysis.
        
        \begin{figure}
            \begin{center}
                \includegraphics[width=1\linewidth,clip]{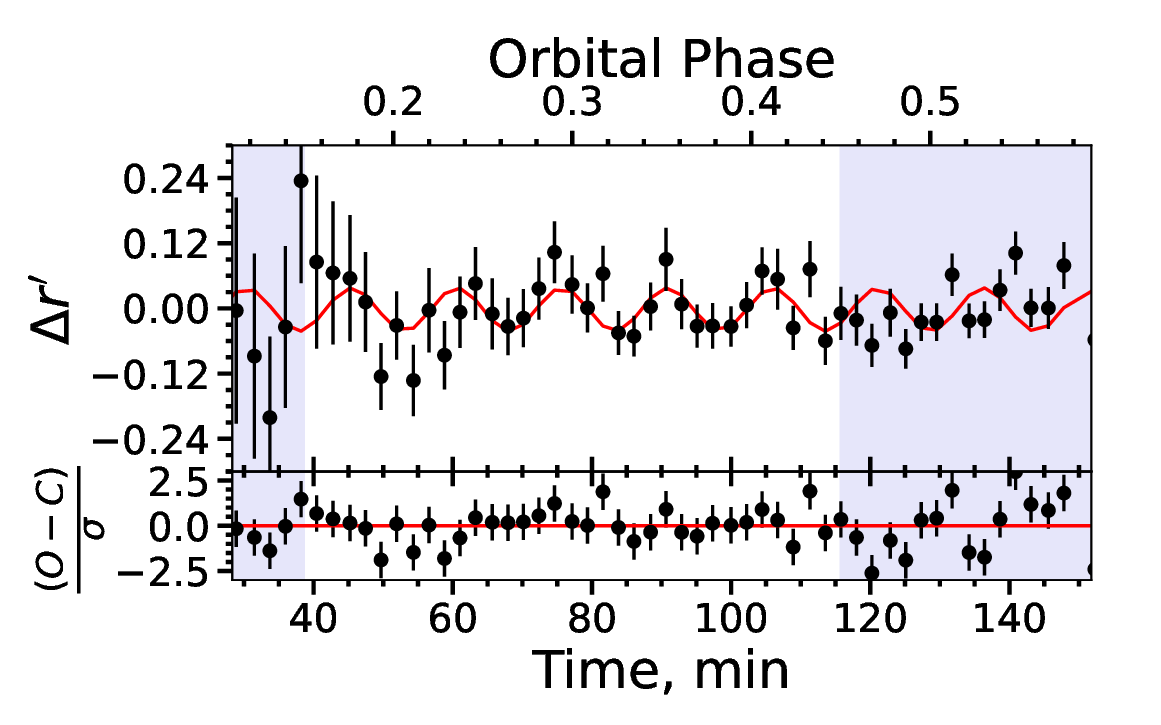}
                \includegraphics[width=1\linewidth]{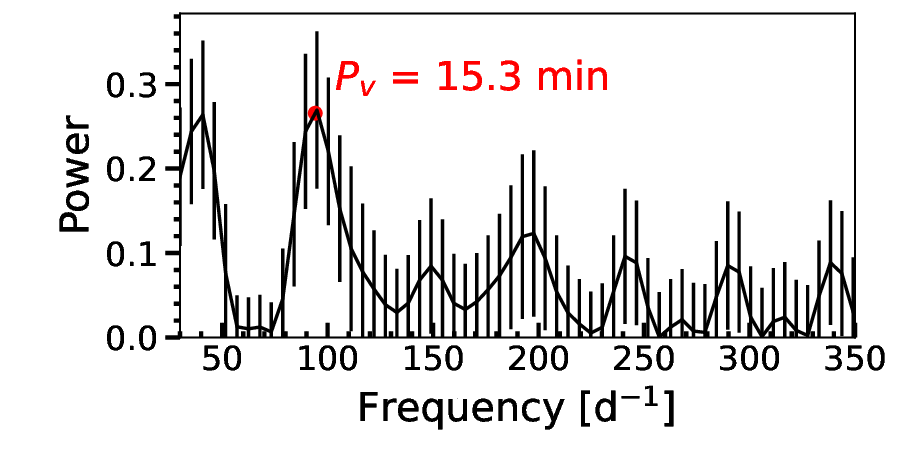}
                \includegraphics[width=1\linewidth]{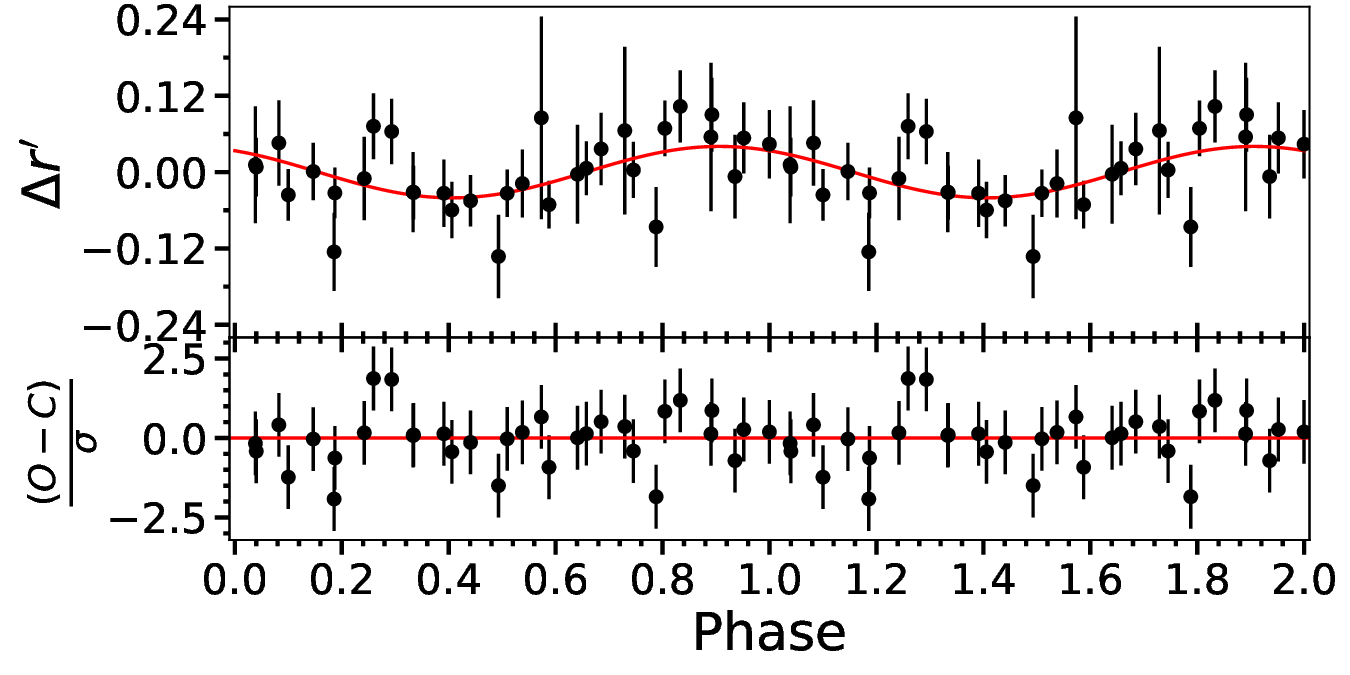}
            \end{center}
            \caption{
                Result of the small-scale brightness variations analysis.
               \textit{Top panel}: J1513 filtered light curve in the $r'$ band (black points with error bars). Points within shaded areas are excluded  from the periodicity search. The red line is the best fit of the data by a sine function with the period of 15.3 min found from the periodicity search. 
               The sub-panel shows fit residuals.
               \textit{Middle panel}: power spectrum of the filtered light curve (the solid line) with its uncertainties (vertical error bars). The largest peak corresponds to the 15.3 min period.
               \textit{Bottom panel}: light curve folded with this period and fitted with the sine function of the same period. The sub-panel shows fit residuals.
            }
            \label{fig:lc1513_fluc}
        \end{figure}
    
        To study the variations, it is natural to filter out the large-scale variations related with the orbital motion. To do that, we fitted the light curve at $\phi \in [0.1,0.6]$ in the $r'$- band with a low order Chebyshev polynomial \citep{Chebyshev_polynomial_b} and subtracted this approximation  from the light curve. The result is presented in the top panel of Fig.~\ref{fig:lc1513_fluc}. The filtered light curve demonstrates an apparent periodic variability for about 75 min between orbital phases of $\sim$0.15 and 0.45. We used the data points from this interval and the Lomb-Scargle periodogram method \citep{lomb1976,scargle1982} provided by the \texttt{astropy} package\footnote{\href{https://docs.astropy.org/en/stable/timeseries/lombscargle.html}{https://docs.astropy.org/en/stable/timeseries/lombscargle.html}} to search for the brightness periodicity. The highest peak in the power spectrum corresponds to the period $P_V \approx 15.3$ min (Fig.~\ref{fig:lc1513_fluc}, middle panel). To estimate the power spectrum uncertainty, we used an equivalent  of the bootstrap method. We simulated 10000 light curves randomly scattered within the brightness intervals defined by $\pm$1$\sigma$ errors of the data, assuming a normal distribution.  We obtained periodograms for each of the simulated curves. Then the periodogram distribution for each frequency bin was fitted with the Gaussian whose maximum position and width values were attributed to the most probable power spectrum value and its error in the bin, respectively. 
        
        The calculated false alarm probability  of the 15.3 min peak is $\approx$0.46, making it   questionable. However, we constructed the power spectra of a non-variable reference star and the window function of the data for the same time range as for \psrn. We did not find any signature of the 15.3-min period there. This indicates  that the 15.3-min peak in \psrn\  is not an artefact.
        
        In addition, we fitted the data with a constant and then with a sine function of the 15.3-min period plus a constant. The latter resulted in $\chi^2=24.1$ for 30 {\sl dof} and the average peak-to-peak amplitude of $\approx$0.08 mag (see Fig.~\ref{fig:lc1513_fluc}, top panel). With the constant model, we obtained a slightly worse but also acceptable fit with  $\chi^2=33.3$ for 32 {\sl dof}.  
        
        The $F$-test applied to compare the two models  gave a probability of $\approx$8$\times$10$^{-3}$ of improvement by chance of the sine fit with respect to the constant one, implying strong  evidence against  the constant model. The light curve folded with the obtained period  shown in the bottom panel of Fig.~\ref{fig:lc1513_fluc}, together with the best sine fit, appear to favour the presence of the periodicity. The remaining part of the filtered light curve shows only stochastic fluctuations of the same amplitude and no signs of the periodicity. Nevertheless, we finally conclude that while the  small-scale variability is clearly visible in  the considered phase interval, its apparent periodicity  has a low statistical  significance of $\approx$2.5$\sigma$ and new observations are needed to confirm it.
        
        
    \section{X-ray data}
        \label{sec:xray}
    
         \begin{figure*}
            \begin{center}
               \includegraphics[width=0.45\textwidth]{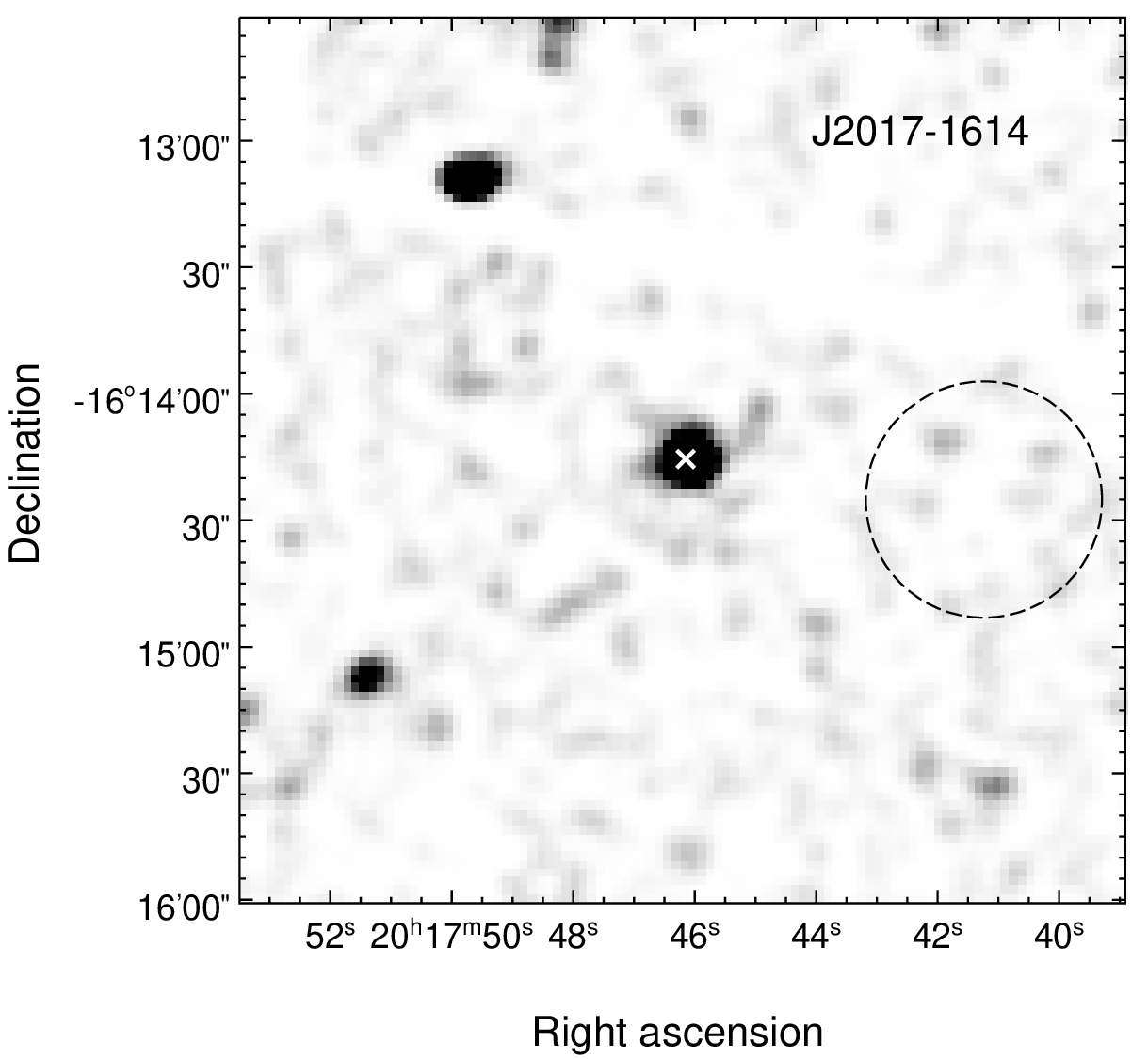} 
               \hspace{0.1cm}
               \includegraphics[width=0.45\textwidth]{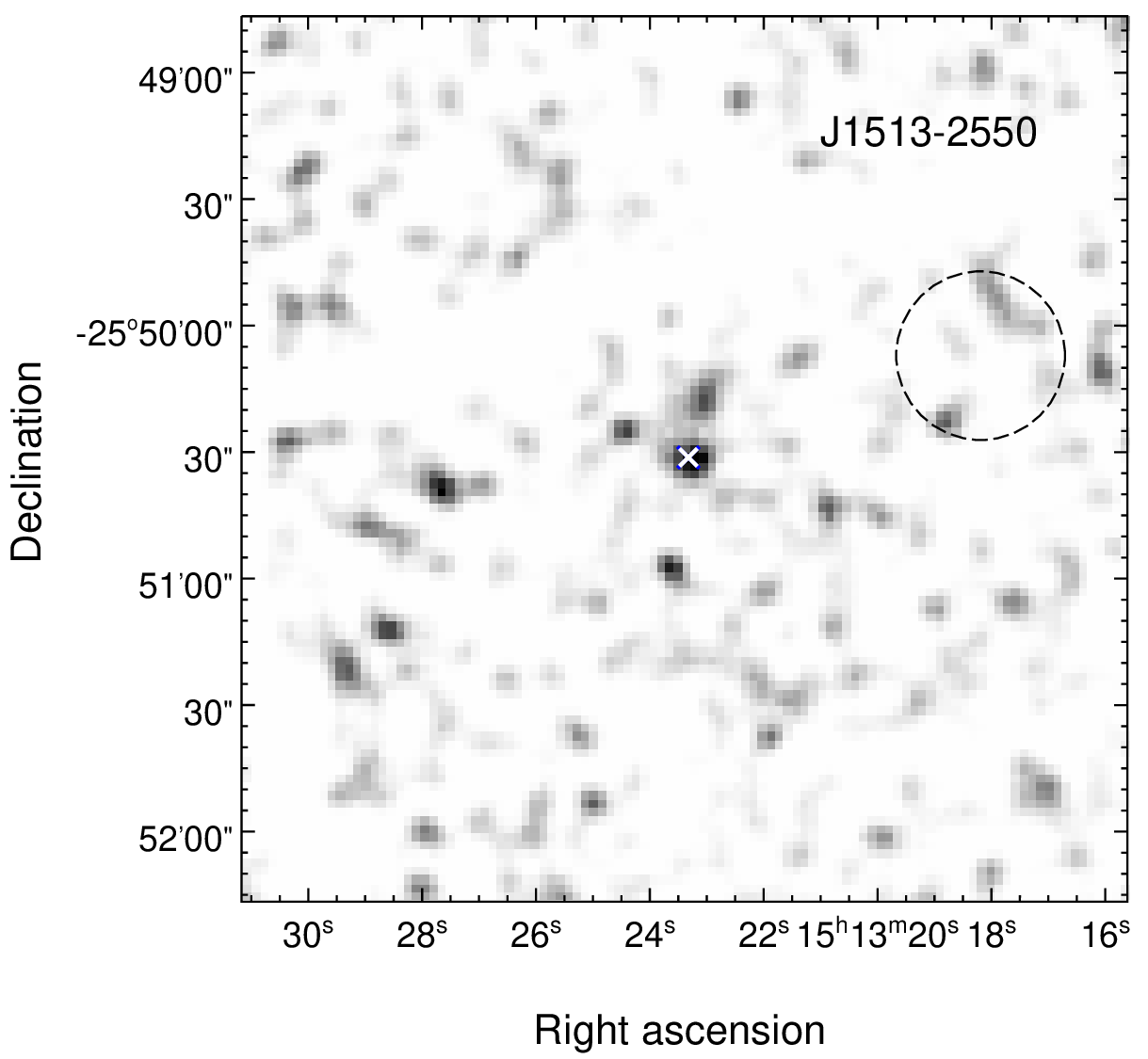}
            \end{center}
            \caption{3\farcm5 $\times$ 3\farcm5 \xmm/PN images of the \psr\ (\textit{left}) and \psrn\ (\textit{right}) fields in the 0.2--2 keV band. The pulsars' radio positions (Table~\ref{tab:pars}) are marked by the `X' symbols. Dashed circles show regions used for the background extraction.}
            \label{fig:xray}
        \end{figure*}
        
        We found X-ray counterparts to \psr\ and \psrn, 4XMM J201746.0$-$161416 and 4XMM J151323.3$-$255029, in the \xmm\ Serendipitous Source Catalogue Data Release 13 \citep[4XMM-DR13;][]{4xmm-dr13}. According to the catalogue, 4XMM J151323.3$-$255029 is very weak, with an absorbed flux $f_X=(5.8\pm2.2)\times10^{-15}$ \flux\ in the 0.2--12 keV band. 4XMM J201746.0$-$161416 is brighter, with  $f_X=(1.8\pm0.3)\times10^{-14}$ \flux. Below, we present an independent analysis of the observations.
        
        For both pulsars, the European Photon Imaging Camera-Metal Oxide Semiconductor (EPIC-MOS) detectors and EPIC-pn (PN hereafter) were operated in the full frame mode with the medium filter. The XMM–Newton Science Analysis Software (\texttt{XMM-SAS}) v.21.0.0 was used for data reduction. We reprocessed the data, applying the \texttt{emproc} and \texttt{epproc} tasks.

        \subsection{\psr}
            The \xmm\ observation of the \psr\ field was performed on October 15 2016 (ObsID 0784770601, PI M. Roberts, duration = 25 ks). Examination of the high-energy light curves extracted from the FoVs of EPIC detectors revealed some strong background flares. We used the \texttt{espfilt} task to filter event lists, which resulted in effective exposure times of 17, 18, and 13 ks for the MOS1, MOS2, and PN instruments, respectively.
            
            The PN image of the pulsar field is presented on the left of Fig.~\ref{fig:xray}. The X-ray source is clearly seen in the pulsar radio position. We extracted its spectra using the 14\asec-radius aperture. The background spectra were extracted from the source-free circle region with a radius of 28\asec. The resulting net counts in the 0.2--10 keV bands are 27, 29, and 72 for MOS1, MOS2, and PN data, respectively. The redistribution matrix and ancillary response files were generated by the \texttt{rmfgen} and \texttt{arfgen} routines.
            
            We grouped the spectra to ensure at least 1 count per energy bin and fitted them simultaneously in the 0.2--10 keV band in the X-Ray Spectral Fitting Package (\texttt{XSPEC}) v.12.13.1 \citep{xspec}. We used the $W$-statistics appropriate for Poisson data with Poisson background and the power law (PL) model that is usually utilised to describe X-ray spectra of spider pulsars \citep{swihart2022,strader2019}. To account for the interstellar absorption, we chose the \texttt{tbabs} model with the \texttt{wilm} abundances \citep{wilms2000}. The equivalent hydrogen column density, $N_{\rm H}$, was fixed at 8$\times$10$^{20}$ cm$^{-2}$. The latter was calculated using the reddening value obtained from the optical light curves modelling (Table~\ref{tab:fit}) and the relation from \citet{foight2016}. As a result, we obtained the photon index $\Gamma=2.7\pm0.2$, the unabsorbed flux in the 0.5--10 keV band $F_X=(1.5\pm0.2)\times10^{-14}$ \flux\ , and $W=80$ for 81 degrees of freedom.

        \subsection{\psrn}
            The \psrn\ field was observed with \xmm\ on September 1 2016 (ObsID 0784770701, PI M. Roberts, duration = 25 ks). We detected only one short flare, which is clearly seen only in the PN light curve. Thus, we filtered the PN data using the count rate threshold of 0.5 counts~s$^{-1}$, which resulted in the effective exposure time of 15 ks.
            
            The \psrn\ X-ray counterpart is very faint (Fig.~\ref{fig:xray}, right) and most prominent in the images from the PN camera, which is more efficient in the soft ($\lesssim$2 keV) band than the MOS detectors. We extracted the source spectra using a 10\asec-radius circle. For the background spectrum, we chose a source-free region with a radius of 20\asec. As a result, we obtained 18 source counts in the 0.2--10 keV band from the PN instrument, while the MOS cameras provided 11 counts.
            
            The spectra were grouped to ensure at least 1 count per energy bin and fitted simultaneously in the 0.2--10 keV band. The reddening for \psrn\ derived from the optical light curves analysis (Table~\ref{tab:fit}) was converted to the equivalent hydrogen column density, $N_{\rm H}=6\times10^{20}$ cm$^{-2}$. This value was fixed during the fitting procedure. The PL model resulted in the photon index $\Gamma=2.0^{+0.3}_{-0.4}$, the unabsorbed flux in the 0.5--10 keV band $F_X=6.2^{+2.2}_{-1.6}\times 10^{-15}$ erg~s$^{-1}$~cm$^{-2}$, and $W=29$ for 44 degrees of freedom.

    \section{Discussion}
        \label{sec:discussion}
    
        We have presented results of optical and X-ray observations of \psr\ and \psrn. 
        For the first time, both pulsars were detected in X-rays and \psrn\ was firmly identified in the optical.
        Below, we discuss their characteristics.

        \subsection{\psr}
            \psr\ belongs to the class of short orbital period systems and  shows a typical optical light curve observed in spider pulsars. Modelling the optical data suggests a small companion mass of $\approx$0.04\msun\ , which confirms the BW nature of \psr. The system NS mass is apparently larger than the canonical one of 1.4~\msun. The companion is cool, with a night-side temperature of $T_n=3000^{+200}_{-100}$ K. The  difference between this and the temperature of a hotter day-side is about a factor of two. These values are typical for BWs \citep[e.g.][]{matasanchez2023}. The distance, $D=2.40^{+0.10}_{-0.05}$ kpc, is larger than the DM distance estimates (Table~\ref{tab:pars}). This is not surprising, as according to \citet{koljonen&linares2023}, DM distances to spider pulsars are systematically lower than those derived by other methods. The $E(B-V)$ value is consistent with the expected  value of $0.11^{+0.01}_{-0.02}$ mag derived for the line-sight of the pulsar and its distance using the 3D Galactic dust map by \citet{dustmap2019}. 
            
            The X-ray spectrum of the system can be described by the PL model with a photon index of 2.7$\pm$0.2. The X-ray luminosity in the 0.5--10 keV band is $\approx10^{31}$ \ergs\ for the distance $D=2.4$ kpc. The flux and corresponding luminosity in the 0.1--100 GeV range is $F_\gamma^{\rm \psr} = 6.5(6)\times10^{-12}$~\flux\ \citep{4fgl-dr3} and $L_\gamma^{\rm \psr}=4.5\times 10^{33}$ \ergs. These values are in agreement with those obtained for other BWs \citep{swihart2022}.
            
            The irradiation efficiency, $\eta=0.6$, calculated for the observed $\dot{E}$ overshoots values typical for BWs \citep[e.g.][]{draghis2019,matasanchez2023}. The $\gamma$-ray efficiency, $\eta_\gamma=L_\gamma/\dot{E}=0.58$, is also sufficiently high. The correction of the pulsar spin-down luminosity for acceleration due to differential Galactic rotation \citep{nice&taylor1995,lynch2018} results in a slightly lower intrinsic value, $\dot{E}_i=7.7\times10^{33}$ \ergs, implying higher efficiencies. However, as was noted above, \psr\ can be a rather massive NS with a larger than canonical moment of inertia, and therefore larger $\dot{E}_i$. Assuming a plausible NS radius, $R_{\rm p}=13$ km \citep[e.g.][]{salmi2022,Vinciguerra2024}, and using the formula from \citet{ravenhall&pethick1994}, we obtained $\dot{E}_i=1.7\times10^{34}$ \ergs\ for the lower bound of the NS mass, $M_{\rm p}=1.8$~\msun\ (Table~\ref{tab:fit}). The corresponding $\eta=0.28$ and $\eta_\gamma=0.26$ are more plausible for BW systems, and they can be even lower if \psr\ is more massive and/or if the beaming factor correction is taken into account \citep[e.g.][]{draghis2019,romani2021}.

        \subsection{\psrn}
            \psrn\ is an example of a spider pulsar with an asymmetric optical light curve. This feature appears in both OAN-SPM and \magel ~ observations performed about three years apart, indicating that this property is persisting. The simple model of direct heating of the companion cannot reproduce the asymmetry. There are a number of other spider systems demonstrating similar behavior; for example, BW J1311$-$3430 \citep{romani2012}, BW J1653$-$0158 \citep{nieder2020}, BW J1810+1744 \citep{romani2021}, RB J2039$-$5617 \citep{clark2021}, and RB J2339$-$0533 \citep{kandel2020}. Different models have been proposed  to explain the asymmetries: cold spots on a companion star surface caused by its magnetic activity \citep{clark2021}, asymmetric heating from the intrabinary shock \citep[IBS;][]{romani&sanchez2016}, IBS particles channelling to the magnetic poles of the companion star \citep{sanchez&romani2017}, and heat redistribution over the companion surface via convection and diffusion \citep{kandel&romani2020, voisin2020}. It is not clear which of those ideas is truly applicable to describe \psrn\ due to the possible presence of puzzling periodic brightness wavering with a peak-to-peak amplitude of $\sim$0.1 mag and a periodicity of $\sim$15 min in the light curve in the $\sim$0.15--0.45 orbital phases (Fig.~\ref{fig:lc1513_fluc}).
            No such variations have been reported so far for other spider systems.
            
            Possible wavering could be attributed to the flaring activity of \psrn\ , which is observed in some other spider systems \citep[e.g.][]{romani2012,Flaring-BW-2022,swihart2022,zyuzin2024}. However, the observed periodicity calls this suggestion into question as flaring is usually a stochastic process. In the case of our data, it is  unclear whether these variations always appear in the same orbital phases.
            
            On the other hand, if the variations are proved in the future, then they could be explained by the sunspot-like oscillations \citep[e.g.][]{sunspot2016,mhd-sunspot-2016} of the companion's hot spot, produced by IBS particles being channelled to the off-axis magnetic pole of the companion star \citep[see fig.~3 in][]{sanchez&romani2017}.

            Though the used symmetric model does not describe the \psrn\ light curves for the whole phase range, it provides preliminary constraints on the system's parameters. The night- and day-side temperatures are close to those obtained for \psr\ (Table~\ref{tab:fit}). The derived distance to \psrn\ is about 2 kpc, which is close to the value provided by the NE2001 model. Taking into account the pulsar proper motion, $\mu=7.2(2)$ mas~yr$^{-1}$, the transverse velocity is $v_t = 68$ km~s$^{-1}$, which is consistent with the velocity distribution for binary pulsars \citep{hobbs}. The best-fitting interstellar absorption for the pulsar is below the maximum value in this direction, $0.12^{+0.02}_{-0.01}$ mag, derived from  the dust map of \citet{dustmap2019}.

            The \psrn\ system has a weak counterpart in X-rays. Its spectrum can be described by the PL model with a photon index of $2.0^{+0.3}_{-0.4}$. The X-ray and $\gamma$-ray luminosities of \psrn\ for a distance of 2 kpc are $L_X^{\rm \psrn}=3.0\times 10^{30}$ \ergs and $L_\gamma^{\rm \psrn}=3.5\times 10^{33}$ \ergs. The latter was calculated from the $\gamma$-ray flux $F_\gamma^{\rm \psrn} = 7.4(6)\times10^{-12}$~\flux\ \citep{4fgl-dr3}. These values are typical of BW systems \citep{swihart2022}.

            The intrinsic  spin-down luminosity of \psrn\ corrected for the Shklovskii effect \citep{shklowskii1970} and acceleration  due to the Galactic differential rotation is $\dot{E}_i=1.7\times10^{35}$ \ergs\ for $D=2$ kpc, $M_{\rm p}=1.7$~\msun\ , and $R_{\rm p}=13$ km. This implies the efficiencies $\eta_\gamma\approx0.02$ and $\eta\approx0.03$. While the $\gamma$-ray efficiency is typical of BWs, the irradiation efficiency is very low. The lowest values provided by \citet{matasanchez2023} for a sample of BWs are 0.068$\pm$0.010 for PSR J2256$-$1024 and 0.09$\pm$0.05 for PSR J0251+2606. \psrn\ has a rather high spin-down luminosity in comparison with other BW pulsars, including that of \psr. However, its orbital separation is also larger, which could imply a similar heating rate of  \psr\ and \psrn\ companions.

    \section{Conclusions}
        \label{sec:conclusion}
        We performed first multi-band time-series optical photometry of two BW pulsars, \psr\ and \psrn.
        This allowed us to firmly identify the latter in the optical.
        Both objects show similar peak-to-peak amplitudes of orbital modulation, $\gtrsim 2$ mag. 
        However, while the \psr\ light curves are symmetric, the \psrn\ demonstrates strong asymmetry and periodic wavering of unknown origin. 
        
        The light curve modelling provides parameters for the \psrn\ and \psr\ companions (masses, radii, temperatures, and Roche-lobe filling factors) comparable with values derived for other BWs \citep[e.g.][]{zharikov2019,swihart2022,matasanchez2023}.
        Inclinations of the systems are about 70\degs\ , which is compatible with observed irregular eclipses of the pulsars' radio emission.
        
        For the first time, \psr\ and \psrn\ are identified in X-rays.
        Their X-ray spectra can be described by the absorbed PL models providing photon indices and luminosities consistent with the values derived for the BW sample.  
        
        New optical observations of \psrn\ with a high
        time resolution are necessary to confirm the wavering and to clarify its nature.
        Detailed multi-band observations would provide information about source colours, which could help us to understand the mechanism causing the light curves asymmetry and to better constrain the system parameters.

    \begin{acknowledgements}
        We thank the anonymous  referee for useful comments.  
        The work is based on observations carried out at
        the 6.5 meter \magel~ Telescopes located at Las Campanas Observatory, Chile,
        the Gran Telescopio Canarias (GTC), installed at the Spanish Observatorio del Roque de los Muchachos of the Instituto de Astrof\'isica de Canarias, in the island of La Palma, 
        the Observatorio Astron\'omico Nacional on the Sierra San Pedro M\'artir (OAN-SPM), Baja California, M\'exico and 
        on observations obtained with XMM–Newton, a ESA science mission with instruments and contributions directly funded by ESA Member States and NASA.
        We thank the daytime and night support staff at the OAN-SPM for facilitating and helping obtain our observations.
        This work has made use of data from the European Space Agency (ESA) mission {\it Gaia} (\url{https://www.cosmos.esa.int/gaia}), processed by the {\it Gaia} Data Processing and Analysis Consortium (DPAC, \url{https://www.cosmos.esa.int/web/gaia/dpac/consortium}). Funding for the DPAC has been provided by national institutions, in particular the institutions participating in the {\it Gaia} Multilateral Agreement.
        The Pan-STARRS1 Surveys (PS1) and the PS1 public science archive have been made possible through contributions by the Institute for Astronomy, the University of Hawaii, the Pan-STARRS Project Office, the Max-Planck Society and its participating institutes, the Max Planck Institute for Astronomy, Heidelberg and the Max Planck Institute for Extraterrestrial Physics, Garching, The Johns Hopkins University, Durham University, the University of Edinburgh, the Queen's University Belfast, the Harvard-Smithsonian Center for Astrophysics, the Las Cumbres Observatory Global Telescope Network Incorporated, the National Central University of Taiwan, the Space Telescope Science Institute, the National Aeronautics and Space Administration under Grant No. NNX08AR22G issued through the Planetary Science Division of the NASA Science Mission Directorate, the National Science Foundation Grant No. AST-1238877, the University of Maryland, Eotvos Lorand University (ELTE), the Los Alamos National Laboratory, and the Gordon and Betty Moore Foundation.
        AYK acknowledges the DGAPA-PAPIIT grant IA105024. 
        SVZ acknowledges PAPIIT grant IN119323.
        DAZ thanks Pirinem School of Theoretical Physics for hospitality. 
        The analysis of the X-ray data by AVK and DAZ was supported by the Russian Science Foundation project 22-12-00048, \url{https://rscf.ru/project/22-12-00048/}.
        Optical data reduction produced by AVB and YAS was supported by the baseline project FFUG-2024-0002 of the Ioffe Institute.
        REM acknowledges support by BASAL Centro de Astrof{\'{i}}sica y Tecnolog{\'{i}}as Afines (CATA) FB210003.
    \end{acknowledgements}
    
    \bibliographystyle{aa}
    \bibliography{ref}
 
\end{document}